\documentclass[epjST]{svjour}
\usepackage{graphicx}
\usepackage{natbib}
\usepackage{enumerate}
\usepackage{amsmath}
\usepackage{upgreek}
\usepackage{color}
\usepackage{ulem}
\usepackage{latexsym}
\usepackage{amsfonts}
\begin{document}
\title{Non-uniformly receding contact line breaks axisymmetric flow patterns}
\author{Hyoungsoo Kim\inst{1}\fnmsep\thanks{\email{hshk@kaist.ac.kr}}\fnmsep\thanks{H. Kim and N. Belmiloud contributed equally to this work.} \and Naser Belmiloud\inst{2}\fnmsep\thanks{H. Kim and N. Belmiloud contributed equally to this work.} \and Paul W. Mertens\inst{3} }
\institute{Department of Mechanical Engineering, KAIST, Daejeon 34141, South Korea \and SCREEN SPE Germany GmbH, Fraunhofer Strasse 7, 85737 Ismaning, Germany \and IMEC, vzw Kapeldreef 75, Heverlee 3001, Belgium}
\abstract{
We investigate the internal flow pattern of an evaporating droplet using tomographic particle image velocimetry (PIV) when the contact line non-uniformly recedes. We observe a three-dimensional azimuthal vortex pair while the contact line non-uniformly recedes and the symmetry-breaking flow field is maintained during the evaporation. Based on the experimental results, we show that the vorticity magnitude of the internal flow is related to the relative contact line motion. Furthermore, to explain how the azimuthal vortex pair flow is created, we develop a theoretical model by taking into account the relation between the contact line motion and evaporating flux. Finally, we show that the theoretical model has a good agreement with experimental results.
}
\maketitle
\section{Introduction}
\begin{figure}[!t]
    \centering
        \includegraphics[trim=0.0cm 0.0cm 0.0cm 0.0cm, clip=true, width=1\textwidth]{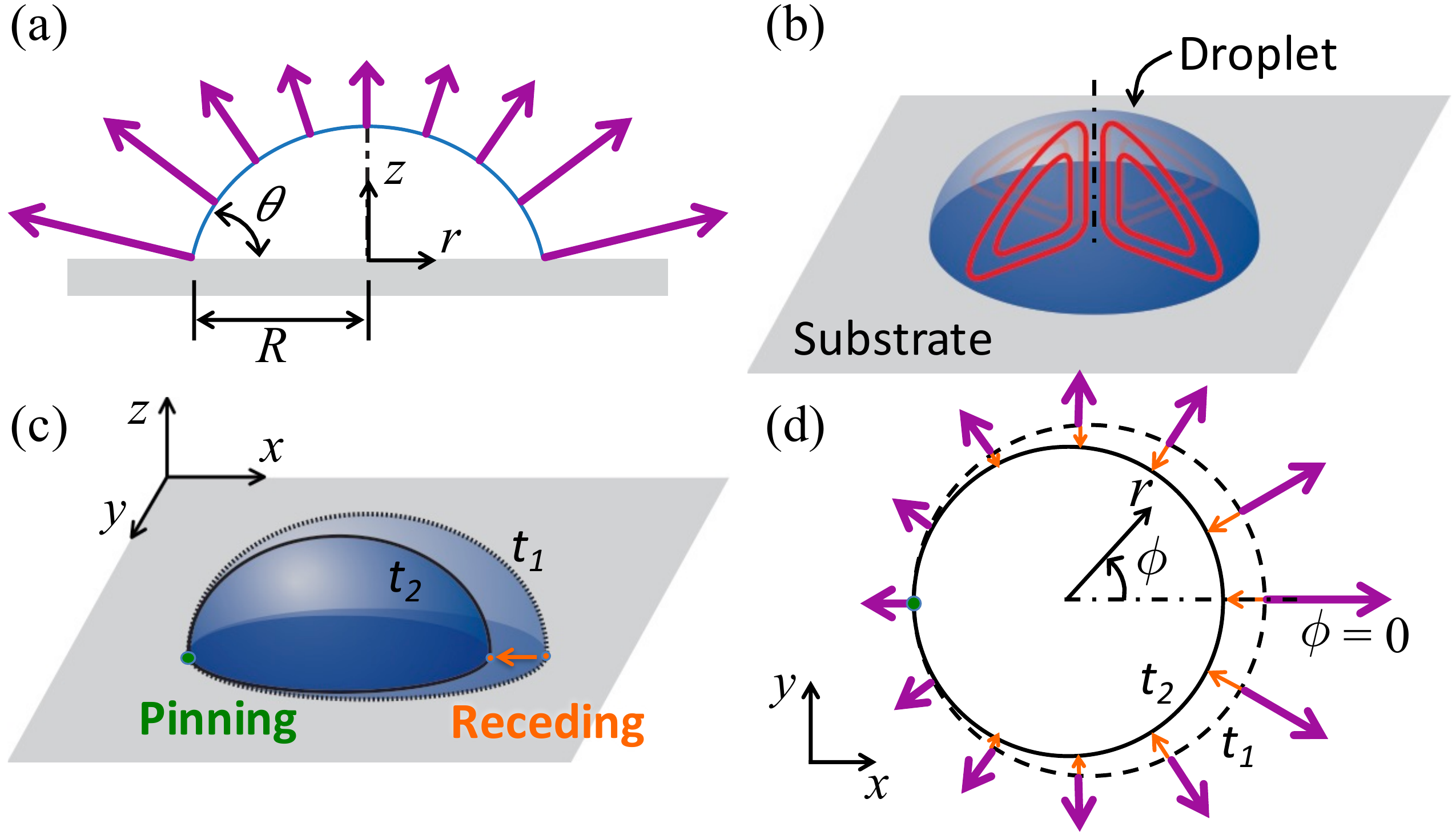}
    \caption{Typical droplet evaporation problem: (a) Schematic of an evaporation rate profile (purple arrows) on the liquid-air interface of a drop on a solid substrate [1]. The length of the purple arrows represents a relative strength of the evaporating flux. (b) Sketch of a radial axisymmetric flow pattern (red-solid lines) where the evaporating flux is axisymmetric. Non-uniformly receding droplet contact line in the azimuthal direction during evaporation: (c) Sketch of a partially receding droplet between $t_{1}$ and $t_{2}$. (d) Schematic of the partially receding contact line in ($r$, $\phi$) coordinate system. The length of the orange arrows represents the relative receding movements.}
    \label{schematic}
\end{figure}

Drying of a liquid droplet on a solid substrate is ubiquitous in ordinary life~\citep{deegan1997capillary,deegan2000contact,kim2016controlled} and industry~\citep{rothschild2004liquid, calvert2001inkjet, kuang2014controllable}. Over the last decade, this phenomenon has been extensively studied because it plays an important role in multidisciplinary applications, including coating and printing methods~\citep{calvert2001inkjet,kuang2014controllable,deegan2000pattern,  park2006control,lee2018uniform}, DNA/RNA microarrays~\citep{lenshof2009acoustic,dugas2005droplet}, and health diagnostics~\citep{martusevich2007morphology,brutin2011pattern}. Since the first study by Deegan $et~al.$~\citep{deegan1997capillary}, most of the theoretical models assumed that the evaporation flux profile is axisymmetric~\citep{deegan1997capillary, hu2002evaporation, hu2005analysis, masoud2009analytical, petsi2008stokes, marin2011order}, which is described as $J(r) = j_{0}[1-(r/R)^2]^{-1/2+\pi/\theta}$ (see Fig.~\ref{schematic}(a)), where $j_{0}$ is the evaporating flux, $R$ is the droplet radius, and $\theta$ is the contact angle~\citep{deegan1997capillary,hu2002evaporation, hu2005analysis,masoud2009analytical, petsi2008stokes, popov2005evaporative}. Then, the resulting flow field is always axisymmetric (see Fig.~\ref{schematic}(b)) if the boundary conditions are uniform along the contact line. In contrast, the axisymmetric results no longer hold if there are non-uniform boundary conditions such as non-uniform surface tension~\citep{park2019control}, non-uniform evaporation rate, non-uniform temperature distribution in the azimuthal direction~\citep{sefiane2013thermal}, which are still ongoing projects.

In-plane thermal fluctuations were experimentally observed in a volatile liquid droplet on a heated surface~\citep{sefiane2013thermal}, a surfactant-added water droplet~\cite{de2014thermocapillary}, and even a sessile water droplet~\citep{karapetsas2012convective}. In the previous literature, it showed that there were thermal waves on the droplet sitting on the heated substrate and they showed that the heat-flux distribution was non-uniform in the azimuthal direction during evaporation where the contact line non-uniformly receded (see Figs. \ref{schematic}(c) and \ref{schematic}(d)). Besides, a non-uniformly receding contact line motion (i.e. stick-and-slip) was observed and studied~\citep{thiele2006depinning,park2012change,bhardwaj2009pattern,orejon2011stick,shanahan1995simple}, which could be due to chemical and physical heterogeneities, but the internal flow field was not fully investigated yet. Furthermore, the complicated three-dimensional flow pattern was numerically investigated where either the droplet contact line had an elliptical shape or after the contact line suddenly dewetted partially~\citep{saenz2015evaporation}. According to our best knowledge, to obtain three-dimensional flow fields inside an evaporating droplet still challenges due to the small volume and the liquid-gas interface.

In this paper, we measure the internal flow pattern when the contact line non-uniformly recedes in the azimuthal direction during evaporation. Here, we use a glass substrate coated with Self Assembled Monolayers (referred to as SAMs). We obtain the three-dimensional azimuthal vortex flow field by means of tomographic PIV, which has been firstly reported in the previous study~\cite{kim2016tomographic}. Here, we further investigate the experimental results. Moreover, to understand this, we develop the theoretical model that is assumed to have a non-uniform evaporating flux model in the azimuthal direction where the apparent contact is constant during evaporation. We assume that a relative contact line receding motion is correlated with a non-uniform evaporating rate along the droplet surface, which could induce a temperature gradient on the droplet interface. Based on this, we conduct an asymptotic analysis of the non-uniformly evaporating droplet problem by solving Stokes equations in the polar coordinate system ($r$, $\phi$). We observe that the asymptotic model shows a good agreement with the experimental result.

\section{Experiments}
\subsection{Substrate preparation} 
We use SAMs in an organic solvent to design transparent substrates with different contact angles~\cite{howarter2006optimization,howarter2007surface}. The molecular self-assembly is mainly promoted by the dominant Van der Waals interaction between the lateral alkyl chains of the SAMs precursors~\cite{schwartz2001mechanisms, zhao1996mechanism,iimura2000study}. SAMs are created by the formation of a siloxane bond between the Si of the silane molecules and the oxygen of the hydroxyl groups on the glass substrate. It is then followed by a slow two-dimensional self-organization of the lateral chains. Initially, the adsorbate molecules form a nematic phase. Then, after a few hours, it forms crystalline structures on the substrate surface. Areas of close-packed molecules nucleate and grow until the surface of a substrate is covered by a single monolayer. The covalent bonds formed during silanization have higher energy and stability with respect to physisorbed films~\cite{fellah2012facilitating}. Furthermore, for dense SAMs, there is no leaching of water inside the substrate (i.e. impermeability of the substrate) and no chemical reaction between substrate and the colloidal solution (i.e. no change of the contact angle or roughness of the substrate). 

In the present study, we prepared the physicochemically homogeneous substrate that SAMs derived from 11-cyanoundecyltrichlorosilane (C$_{12}$H$_{22}$Cl$_{3}$ NSi) are deposited on 3$\times$3 cm$^{2}$ glass substrates pre-cleaned by 15 min exposure to UV-ozone treatment in a Jelight UVO Cleaner. The cleaned glass substrates are immersed for 30 min in a 5$\times$10$^{-3}$ M solution of 11-cyanoundecyltrichlorosilane using toluene as a solvent. The resulting SAM is rinsed with toluene, ethanol, and deionized water and finally dried with nitrogen flow. The substrate surface presents cyano groups terminations, which are chemically inert with respect to water and colloids (in this study, fluorescent particles for the particle image velocimetry (PIV)).

The measured water contact angle on the SAMs substrate is 50 $\pm$ 3$^{\circ}$ according to a standard side-view shadowgraphy measurement. In this experiment, the initial volume of the sessile droplet is 1.5 $\upmu$l. The initial wetting area has a diameter of 3~mm. The initial droplet height is 0.7~mm at the center. The substrate sample is stored under ultra-clean dry-nitrogen ambient before experiments.

\subsection{Tomographic particle image velocimetry} 
\begin{figure*}
    \centering
        \includegraphics[trim=0cm 0cm 0cm 0cm, clip=true, width=1\textwidth]{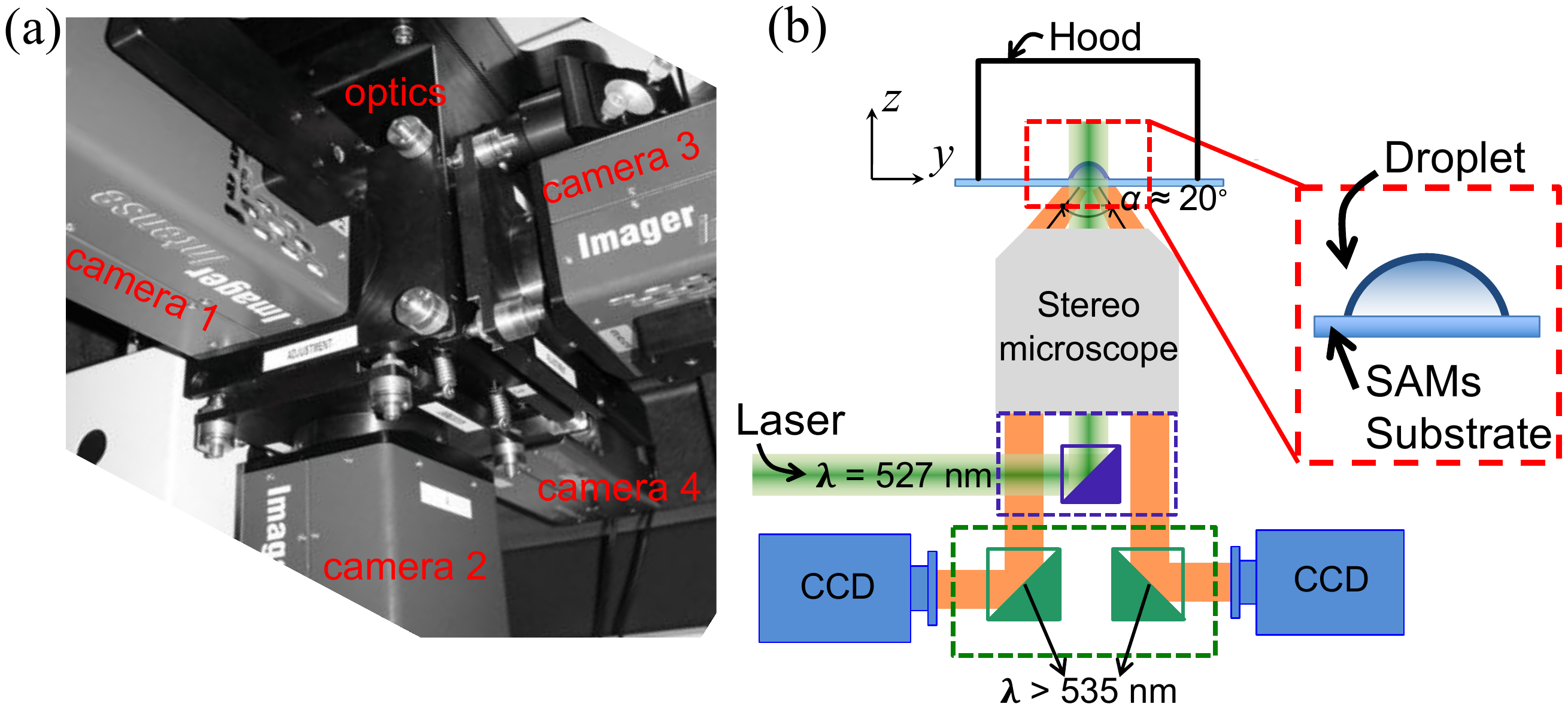}
    \caption{Tomographic particle image velocimetry (PIV) experimental setup for internal flow visualization of a evaporating droplet on a substrate. (a) Picture of tomographic PIV setup with four cameras mounted under a custom-made microscope. (b) Cross-sectional schematic of tomographic PIV system. The angle between the offset optical axes of cameras is approximately $\alpha$ $\approx$ 20$^{\circ}$. The hood covers the measurement area to isolate the external convection effects and to keep a constant temperature and humidity.}
     \label{experiments}
\end{figure*}
A schematic of the experimental setup is given in Fig.~\ref{experiments}. 
To observe the flow field inside a droplet, the entire droplet needs to be accessed without minimal optical distortion for instance caused by reflection or refraction at the liquid-air interface. In this sense, the measurement system is mounted on an inverted custom-made microscope (see Fig.~\ref{experiments}). To perform tomographic PIV to measure the three-dimensional flow, we installed 4 CCD cameras (LaVision Imager intense, 1376 $\times$ 1040 pixels, 12-bit dynamic range) underneath an evaporating droplet. The cameras are mounted with an inverted custom-made microscope (Carl Zeiss Neolumar S1.5$\times$) is equipped with optical filters set (see schematic of the tomographic PIV system in Fig.~\ref{experiments}).

For the measurement of the volumetric flow field inside the droplet, the working fluid (distilled water) is seeded with 1.0 $\upmu$m hydrophilic tracer particles (Duke Scientific Corp.) that are composed of polystyrene and fluorescent (Rhodamine-B). The particle volume fraction is 1.0 \% in aqueous suspension of 1 $\upmu$l and then the particles are diluted in 30 ml distilled water, so that the final particle concentration is approximately 3.3 $\times$ 10$^{-5}$~\% by volume. We checked that the particle solution is almost surfactant-free from a radially outward capillary flow when the contact line is totally pinned during droplet evaporation (see Supplementary Movie 1). In this experiment, the Stokes number, defined as $\mathrm{St}=\tau_p/\tau_f$, where $\tau_p$ is the particle response time and $\tau_f$ is the time scale of fluid motion, is of the order of 10$^{-10}$ because $\tau_p$ (= $\frac{1}{18}\rho_{p}$$d_{p}^{2}$/$\mu$) $\sim$ 10$^{-8}$ s and $\tau_f$ (= $R$/$U$) $\sim$ 10$^{2}$ s, where $\rho_{p}$ is the density of the particle (1050 kg/m$^{3}$ for polystyrene), $d_{p}$ the diameter of the particle, $R$ the droplet radius, and $\mu$ the dynamic viscosity of water~\cite{adrian2010particle}. $U$ is the average flow speed, $\mathcal{O}$(10 $\upmu$m/s), as obtained from the tomographic PIV measurement. Furthermore, the sedimentation effect due to the particle density is almost negligible because the sedimentation particle speed $U_{s}$ $(=g(\rho_{p}-\rho)d_{p}^2/(18\mu))$ is much smaller than the flow speed, i.e. $U_{s}/U$ $\sim$ 10$^{-3}$. Therefore, we believe that the particles well follow the internal flow. 

A frequency-doubled Nd:YLF laser (Pegasus Laser 10 mJ, 527 nm) is used as a light source, which illuminates the entire droplet.
The particles are only illuminated by the laser and there are some reflection and refraction due to the liquid interface, which are improved by a tomographic reconstruction technique. 
To reduce these light distortions, the laser is illuminated from below and all the cameras are installed from below as well (see Fig.~\ref{experiments}).
The frequency of the flashing laser is 1 Hz to obtain the sequence PIV images and the laser exposure time is very short, e.g. shorter than $\mathcal{O}$(1 $\upmu$s).
Furthermore, when the contact line of the droplet totally pins, we observe a capillary flow driven by the evaporation~(see Supplementary Movie 1), which reads that there is no significant laser heating effect.
The camera frames and the lasers are synchronized using a commercial LaVision Programmable Timing Unit (PTU). 

Due to the volume illumination, the most of the particles are scattered. Therefore, it is required to perform a proper image processing to minimize the background noise. First, a min-max filter is applied to normalize the image contrast~\cite{adrian2010particle}. Then, we use a sliding minimum filter to subtract background illumination and a 3 $\times$ 3 Gaussian smooth to reduce image noises on the particles. We perform a 3D calibration with volume self-calibration~\cite{wieneke2008volume}, which reduces the calibration errors to 0.077 pixels corresponding to about 0.3 $\upmu$m. The particle image displacement within a chosen interrogation volume (16 $\times$ 16 $\times$ 16 voxels) with 50$\%$ overlap is obtained by the 3D cross-correlation of the reconstructed particle distribution at the two exposures. The whole pre- and post-processing and calibration procedure are performed by a commercial software (Davis 8.08, LaVision GmbH). In this PIV measurement, the divergence RMS error ($\sigma_{\Delta x}$)~\cite{adrian2010particle} is 0.12 pixels, which is less than 1 $\upmu$m in the actual system. This RMS error is consistent with a typical measurement uncertainty reported for tomographic PIV~\cite{kim2013comparison,elsinga2008tomographic}. The further experimental details and accuracy of the current measurement techniques are described in the preceding studies~\cite{kim2013comparison,kim2011full}.

\subsection{Experimental results}
We observe an internal flow field of a water droplet on a SAMs substrate during evaporation. To maintain a constant temperature ($T_{0}$ = 295 K) and relative humidity (R$_{H}$ = 40$\%$) during experiments, the water droplet was kept inside a hood (see Fig.~\ref{experiments}(b)). A 1.5 $\upmu$l distilled water (DI) water was deposited on top of a SAMs substrate, which is a physicochemically homogeneous substrate. To check the reproducibility, we performed an identical experiment for several times. In this study, a dense single monolayer was uniformly coated as the SAMs layer. The DI water has dynamic viscosity $\mu$ = 10$^{-3}$ Pa$\cdot$s, surface tension $\gamma$ = 0.072 N$\cdot$m$^{-1}$, and density is $\rho$ = 998 kg$\cdot$m$^{-3}$. The surface tension is relatively dominant compared to the inertia and the viscous forces where the Weber and capillary numbers are much smaller than unity, i.e. $We$ = ($\rho U^{2} R$)/$\gamma$ $\ll$ 1 and $Ca$ = $\mu U$/$\gamma$ $\ll$ 1, where the droplet radius $R$ = 1.5 mm and the averaged flow velocity $U$ (obtained from the tomographic PIV) is $U$ = 10 $\upmu$m/s. The initial Bond number ($Bo$) is ($\rho g R^{2}$)/$\gamma$ = 0.3 where gravity $g$ = 9.8 m/s$^{2}$. As the droplet evaporates, $Bo$ will be decreased further. Furthermore, the Reynolds number $Re$ = $\rho$$U$$R$/$\mu$ $\ll$ 1. Thus, gravitational and inertia effects are negligible here.

\begin{figure}
    \centering
        \includegraphics[trim=0.1cm 0.15cm 0cm 0.15cm, clip=true, width=0.85\textwidth]{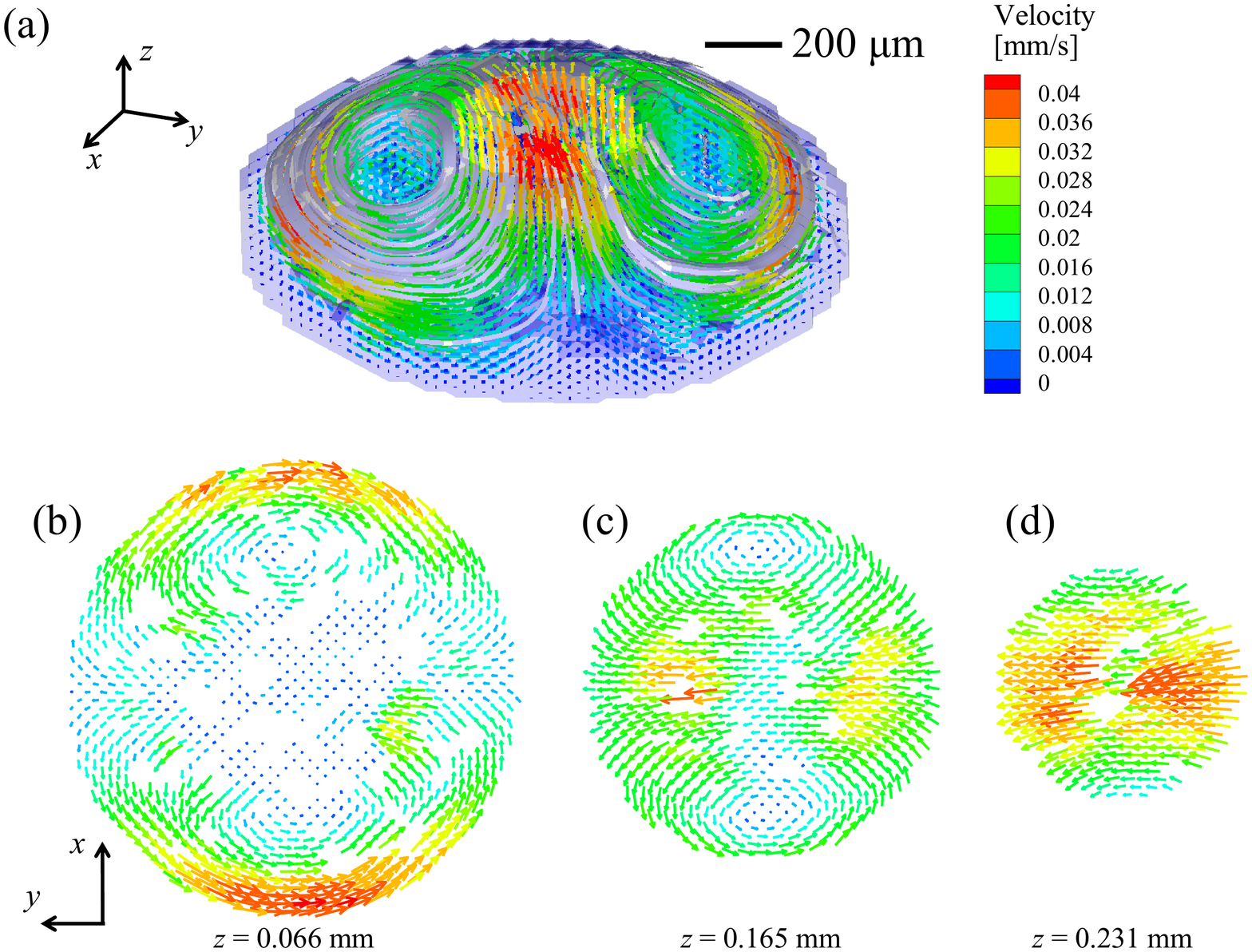}
    \caption{Tomographic PIV result. (a) Full 3D flow pattern inside evaporating droplet at $t$ = 760 (s). Streamlines (white solid-bold lines) and velocity vectors (colors) in cartesian coordinates where the colors represent the velocity magnitude. The bluish contour indicates the liquid-air interface, which is reconstructed by tomography. In-plane flow patterns extracted from (a) at a different height, i.e. (b) $z$ = 0.066 mm, (c) 0.165 mm, and (d) 0.231 mm.}
    \label{3Dflow}
\end{figure}

To date, although several experimental and theoretical studies are performed~\citep{thiele2006depinning,park2012change,bhardwaj2009pattern,orejon2011stick}, the internal flow field is not investigated, except for our previous study\cite{kim2016tomographic}. Here, using tomographic PIV,  we obtain a three-dimensional azimuthal vortex pair inside an drying droplet on a SAMs substrate where the contact line partially recedes in time and space, as shown in Fig.~\ref{3Dflow} and Supplementary Movie 2. An instantaneous flow field at $t$ = 760 (s) after depositing the drop on a substrate is shown as Fig.~\ref{3Dflow}. The typical in-plane flow patterns show a pair of votices at different heights. However, at the top apex, to satisfy the mass conservation, there is an opposite flow pattern. The maximum flow velocity is about 40 $\upmu$m/s located at the liquid interface, which is almost 10 times faster than a maximum speed of the evaporatively-driven capillary flow~\citep{marin2011order}. The flow pattern is not axisymmetric and this vortical flow is maintained until totally dried out.

\begin{figure}
  \centering
      \includegraphics[trim=0cm 0.16cm 0cm 0.15cm, clip=true, width=0.75\textwidth]{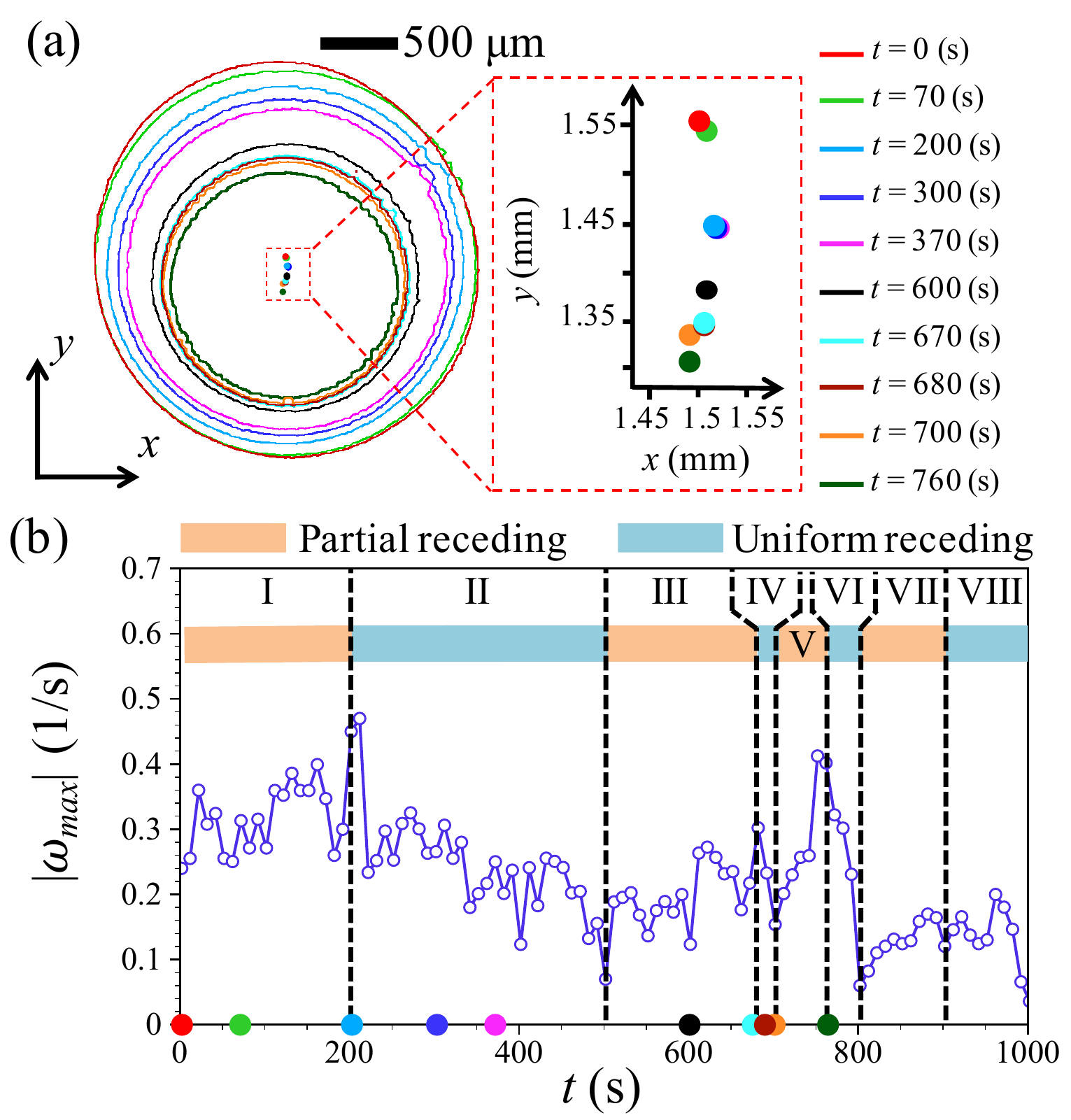}
  \caption{(a) Time evolution of contact lines and center positions of an evaporating droplet. (b) Evolution of the maximum vorticity magnitude during evaporation. In regimes, I, III, V, and VII, the contact lines typically non-uniformly recede. In regimes, II, IV, and VI, the contact lines almost uniformly recede. VIII is the end of drying. The relative speed of the center position is approximately at most $\mathcal{O}$(1 $\upmu$m/s). The different color circles indicate a different time shown in (a) and (b), respectively.}
  \label{flow_evolution}
\end{figure}
To further investigate the flow evolution, we observe and monitor the flow field nearly close to the substrate (at $z$ = 50 $\upmu$m) where the relative contact line receding effect is most dominant (see Fig.~\ref{3Dflow}). We compute the wall-normal vorticity $\omega$ = $\left(\frac{\partial u_y}{\partial x} - \frac{\partial u_x}{\partial y}\right)$ for in-plane velocity $(u_x, u_y)$. We show that the maximum vorticity magnitude $\omega_{max}$ is associated with the receding pattern of the contact line (see Fig.~\ref{flow_evolution}). By monitoring the change of the droplet center and contour, we can estimate the relative contact line receding motion. As shown in Fig.~\ref{flow_evolution}, in I, III, V, and VII regimes, the vorticity magnitude increases as the contact line recedes non-uniformly. On the other hand, in II, IV, and VI regimes, contact lines relatively uniformly recede everywhere (see the blue and magenta lines (or center positions) of Fig.~\ref{flow_evolution}(a) where we obtaine the typical contact line dewetting speed is $\mathcal{O}$(1 $\upmu$m/s)) and the vorticity magnitude decreases in II, IV, and VI regimes. 

\section{Analytical model}
\subsection{Geometric analysis on non-uniformly receding contact line}
\begin{figure}
    \centering
        \includegraphics[trim= 0cm 0cm 0cm 0cm, clip=true, width=0.5\textwidth]{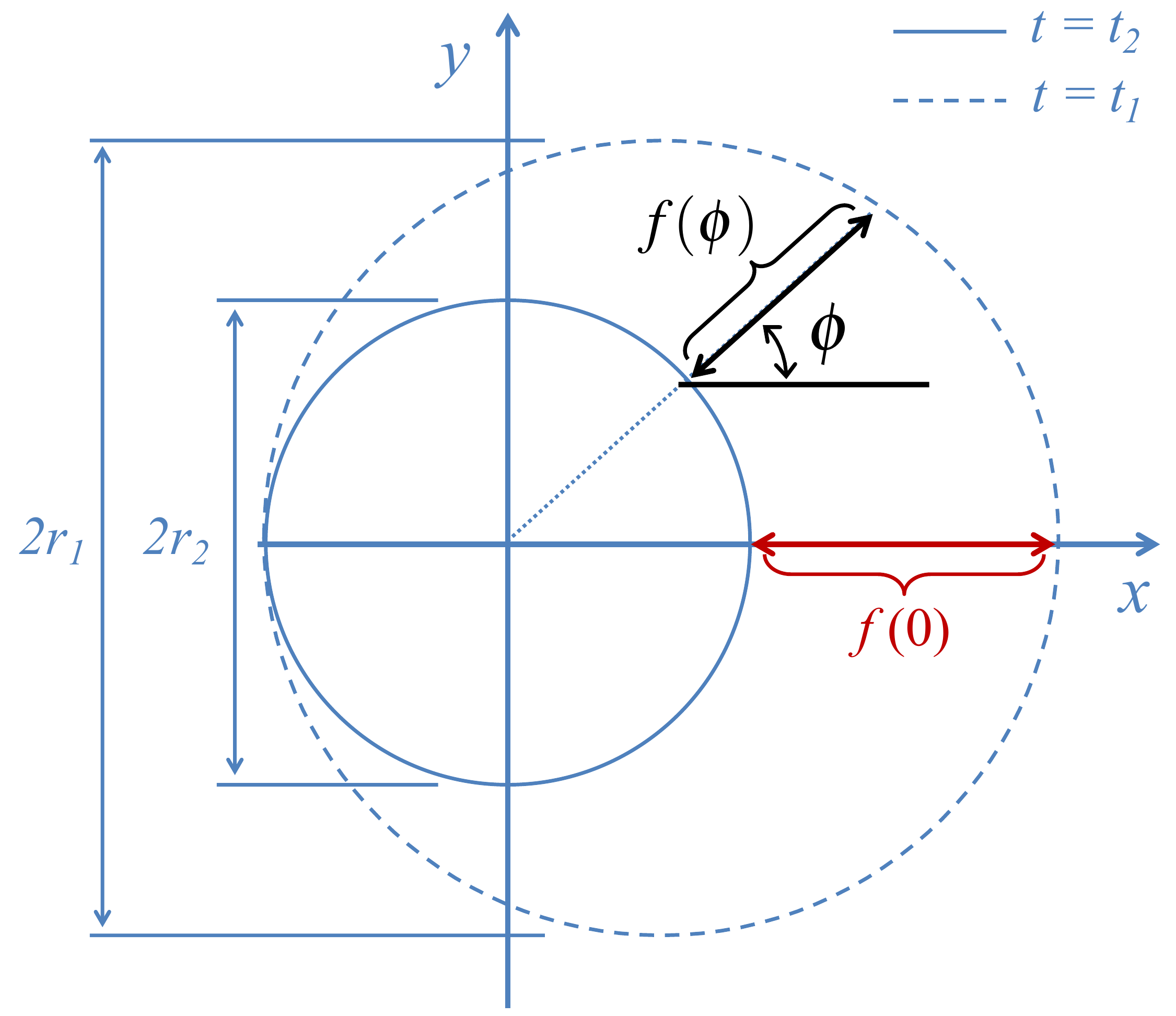}
    \caption{Sketch of a relative receding contact line motion between two different times, i.e. $t_2$ $>$ $t_1$. $f(\phi)$ is the relative distance along the azimuthal direction between two circles. $f(0)$ is the maximum distance between two circles. $f(0) = 2|r_{1}-r_{2}|$.}
    \label{geometric}
\end{figure}
\begin{figure}
    \centering
        \includegraphics[trim= 0cm 0cm 0cm 0cm, clip=true, width=0.5\textwidth]{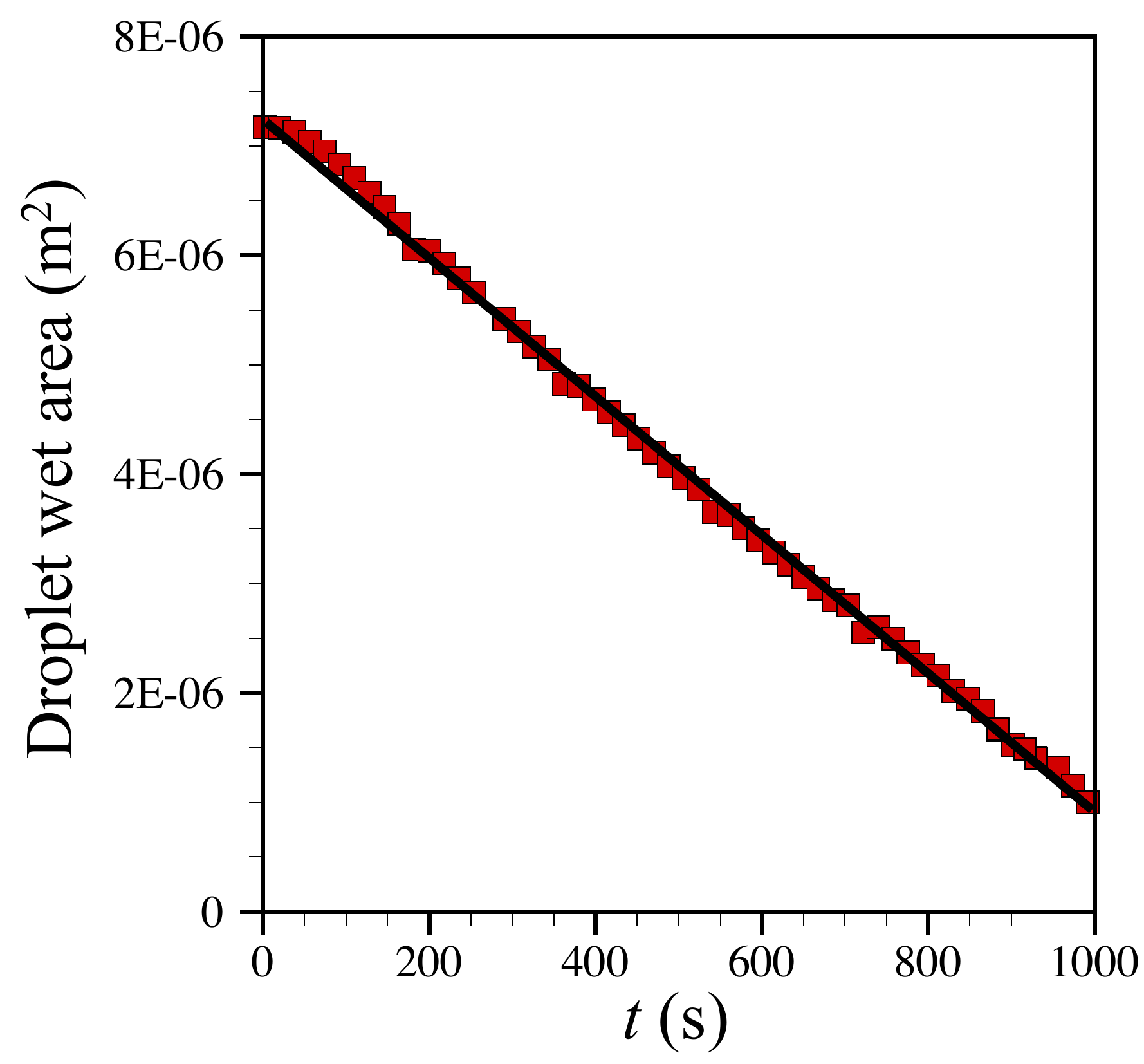}
    \caption{Time evolution of the droplet wet area. The red square symbol is obtained from the experimental observation. The solid line represents a linear fit to the results, i.e. the droplet wet area (m$^{2}$) = (-6.45 $\times$ 10$^{-9}$)$t$ + 7.15 $\times$ 10$^{-6}$.}
    \label{area_profile}
\end{figure}

We consider a non-uniformly receding contact line motion as sketched in Fig.~\ref{geometric}. We assume that one side relatively quickly recedes compared to the opposite side only recedes during evaporation. We obtain the relative contact line shrinkage profile along the azimuthal direction between two different times. From the geometric analysis, the profile is described as:
\begin{equation}
f(\phi) = -\alpha + (1-\alpha)\mbox{cos}\phi + \sqrt{1-(1-\alpha)^{2}\mbox{sin}^2\phi},
\label{geo1}
\end{equation}
where $\alpha$ is the radius ratio between two different times, i.e. $r_{2}/r_{1}$ where $r_1$ and $r_2$ are the radius of a droplet at $t = t_1$ and $t = t_2$, respectively. The time evolution profile of the droplet area is obtained from a fit to the experimental observation as shown in Fig.~\ref{area_profile}, i.e. $r(t)$ = $\pi^{-1/2}$$\sqrt{(-6.45 \times 10^{-9})t + 7.15 \times 10^{-6}}$. 

To obtain the Fourier series of the function $f(\phi)$, we first obtain the Fourier series of the function $g(\phi)$ = $\sqrt{1-(1-\alpha)^{2}\mbox{sin}^2\phi}$ of Eq.~\ref{geo1}. 
The function $g(\phi)$ can be expressed as a Taylor series:
\begin{equation}\label{function_g1}
g(\phi) = \sqrt{1 - (1-\alpha)^2\mbox{sin}^2\phi} = \sum^{\infty}_{i = 0} \binom{1/2}{i}(-1)^{i} [(1-\alpha)\mbox{sin}\phi]^{2i},
\end{equation}
where $\sqrt{1 - (kx)^2} = \sum^{\infty}_{i = 0} \binom{1/2}{i}(-1)^{i} [kx]^{2i}$ and $\binom{m}{n}$ is the binomial coefficient, i.e. $m!/(n!(n-m)!)$.
Furthermore, the Fourier series of sin$^2$($\phi$) is obtained as
\begin{equation}\label{sin2}
\mbox{sin}^{2i}\phi = \frac{1}{2^{2i}}\binom{2i}{i}+\frac{(-1)^i}{2^{2i-1}}\sum^{i-1}_{j = 0}(-1)^j\binom{2i}{j}\mbox{cos}(2(i-j)\phi).
\end{equation}
From Eq.~\ref{function_g1} and~\ref{sin2}, the Fourier series of $g(\phi)$ reads
\begin{equation}\label{function_g2}
\begin{split}
g(\phi) & = \underbrace{\sum^{\infty}_{i = 0} (-1)^{i} \binom{1/2}{i} \left(\frac{(1-\alpha)}{2}\right)^{2i}\binom{2i}{i}}_\text{g$_{1}$($\phi$)} \\
& + \underbrace{\sum^{\infty}_{i = 0} \binom{1/2}{i} \left(\frac{(1-\alpha)^{2i}}{2^{2i-1}}\right)\sum^{i = 1}_{j = 0} (-1)^{j}\binom{2i}{j}\mbox{cos}\left[2(i-j)\phi\right]}_\text{g$_{2}$($\phi$)}.
\end{split}
\end{equation}

As (1-$\alpha$) $<$ 1, $g_{1}(\phi)$ will converge to
\begin{equation}
g_{1}(\phi) = \sum^{\infty}_{i = 0} (-1)^{i} \binom{1/2}{i} \left(\frac{(1-\alpha)}{2}\right)^{2i}\binom{2i}{i} = \frac{2E[(1-\alpha)^2]}{\pi},
\end{equation}
where $E$ is the complete elliptic integral of the second kind, i.e. $E[(1-\alpha)^2]$ 
$=\int_{0}^{\pi/2}\sqrt{1-(1-\alpha)^4\mbox{sin}^2\phi}~d\phi$.

For $g_2(\phi)$, let $n$ = $i-j$,
\begin{equation}\label{g2}
\begin{split}
g_{2}(\phi) & = \sum^{\infty}_{i = 0} \binom{1/2}{i} \left(\frac{(1-\alpha)}{2}\right)^{2i-1}\sum^{i-1}_{j = 0}\binom{2i}{j}\mbox{cos}\left[2(i-j)\phi\right]\\
& = \sum^{\infty}_{n = 0}\sum^{\infty}_{j = 0}\binom{1/2}{n+j}\frac{(1-\alpha)^{2(n+j)}}{2^{2(n+j)-1}}(-1)^{j}\binom{2(n+j)}{j}\mbox{cos}(2n\phi),
\end{split}
\end{equation}

The re-arranged function $g_{2}(\phi)$ reads to 
\begin{equation}\label{g2_2}
g_{2}(\phi) = \sum^{\infty}_{n = 0}\binom{1/2}{n} \frac{(1-\alpha)^{2n}}{2^{2n-1}}\mbox{cos}(2n\phi) F_{1}\left[ -\frac{1}{2} + n; \frac{1}{2} + n; 1 + 2n; (1-\alpha)^{2} \right],
\end{equation}
where $F_{1}$ is the Gauss's hypergeometric function, i.e. $F_{1}$ = $\sum^{\infty}_{i = 0}[( (n-1/2)_{i}(n+1/2)_{i}$ $k^{2i})/( i! (2n+1)_{i})]$.
\\

Finally, the Fourier series of the function $f(\phi)$ reads
\begin{equation}
\begin{split}
f(\phi) = & \frac{2E(1-\alpha)^2}{\pi}-\alpha+(1-2\alpha)\mbox{cos}\phi~+ \\
& \sum^{\infty}_{n=1}\frac{(1-2\alpha)^{2n}}{2^{2n-1}}\binom{1/2}{n}\mbox{cos}(2n\phi) F_{1}\left[ -\frac{1}{2} + n; \frac{1}{2} + n; 1 + 2n; (1-\alpha)^{2} \right],
\end{split}
\end{equation}
where the function $f(\phi)$ represents the relative dewetting motion. We postulate that this realtive motion might induce a shear stress along the contact line. Eventually, the vortical flow patterns are observed when the contact line irregularly dewetted.

In this study, we assume that the relative dewetting motion possibly comes from a non-uniform evaporating flux along the contact line. Namely, we model that at the pinned site the evaporating flux is lower than the partially receding site. This non-uniform evaporating rate could be associated with a temperature profile along the contact line. As a result, we assume that thermal-Marangoni effects could occur along the azimuthal direction. Then, we further assume that the contact line mobility is proportional to the evaporation flux, as sketched in Fig.~\ref{schematic}(c,d). Then, the mass flux of vapor from the evaporating droplet can be described as
\begin{equation}\label{flux-model}
J(R) \cong \frac{Dc_{v}(1 - R_{H})}{R}f(\phi) = j_{0}f(\phi) \quad \mbox{at $r$ = $R$,}
\end{equation}
where $D$ is the diffusion coefficient; $c_{v}$, the saturated concentration of water in the air; and $R_{H}$, the relative humidity. Here, the analytical model has a partially receding contact line, i.e. receding at $\phi = 0$ and pinning at $\phi = \pi$. In this study, we only consider non-uniform evaporation rate effect in the azimuthal direction. Hence, we assume that the contact angle is 90$^{\circ}$ ($\theta$ = $\pi$/2) and then the evaporating flux along the droplet height is assumed to be uniform~\citep{hu2002evaporation,hu2005analysis}. $f(\phi)$ describes a non-uniform evaporation flux profile in the azimuthal direction, i.e. $f(\phi)$ = $-\alpha + (1-\alpha)\mbox{cos}\phi + \sqrt{1-(1-\alpha)^{2}\mbox{sin}^2\phi}$ where 0 $<$ $\phi$ $\leq$ $2\pi$. This function is obtained based on a relative contact line shrinkage between two different times, which is obtained from the geometrical analysis of a partially receding contact line (see the examples in Fig.~\ref{schematic}(c,d) and Fig.~\ref{geometric}). $\alpha$ is the change rate of droplet radii, which is experimentally obtained from Fig~\ref{area_profile}, where $\alpha$ = $r$($t_2$)$/$$r$($t_1$) and $r(t)$ = $\pi^{-1/2}$$\sqrt{(-6.45 \times 10^{-9})t + 7.15 \times 10^{-6}}$.

\begin{figure*}
    \centering
        \includegraphics[trim= 0cm 0cm 0cm 0cm, clip=true, width=1\textwidth]{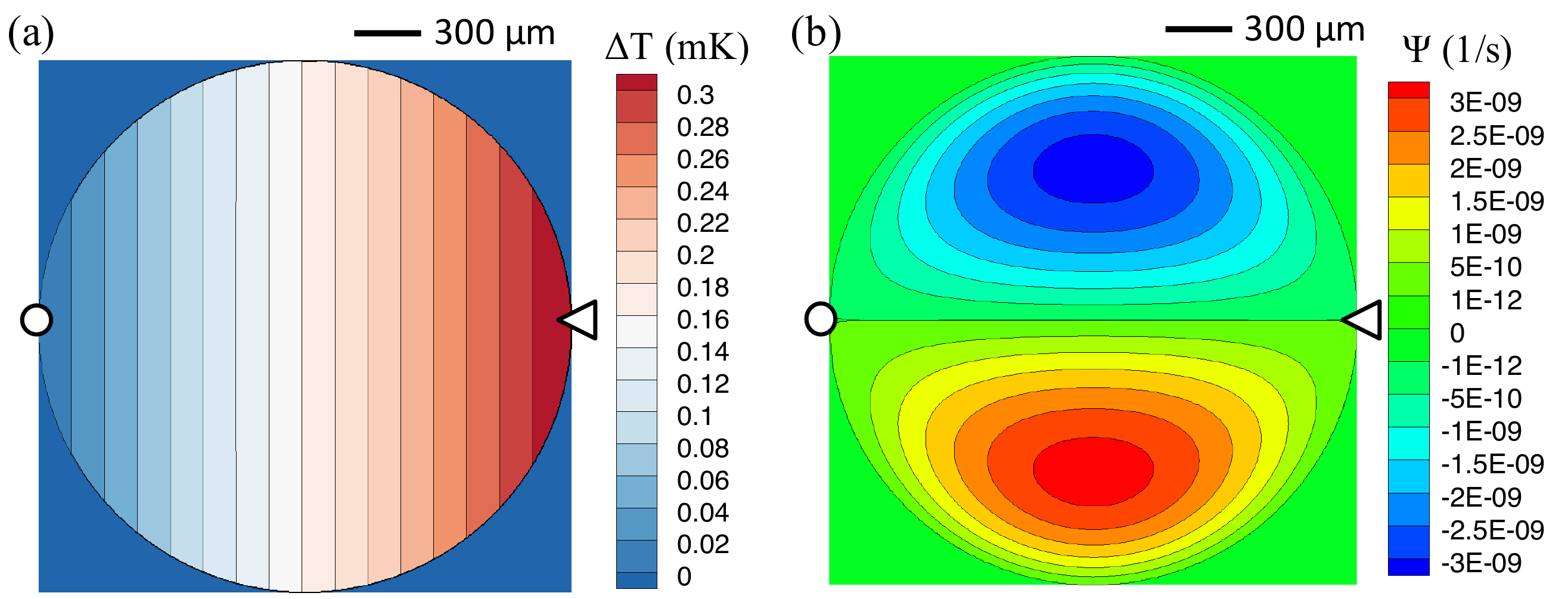}
    \caption{Analytical solutions for (a) temperature distribution and (b) streamfunctions at $t$ = 186 s where the time difference $\Delta t$ is 1 s. $\circ$ and $\triangleleft$ are the relative pinning site and receding site, respectively.}
    \label{analytic_results}
\end{figure*}

If we can assume that the evaporation flux is no longer uniform along the azimuthal direction, the temperature distribution should be also non-uniform. In this problem, the diffusion (evaporation) is predominant and there are no external forces. Biot number $\ll$ 1 ($Bi$ = $hR$/$k_{w}$; $h$ the heat transfer coefficient in the air and $k_{w}$ the thermal conductivity of water) and therefore we neglect heat conduction and convection in the air~\citep{lorenzini2013thermal}. The interfacial energy balance at the liquid-gas interface is 
\begin{equation}\label{conduction}
-k_{w}\mathbf{n}\cdot\nabla T = \Delta H_{v}j_{0}f(\phi) \quad \mbox{at $r$ = $R$},
\end{equation}
where $\textbf{n}$ is the unit normal vector and $\Delta H_{v}$ is the specific latent heat of evaporation. Substitute Eq.~(\ref{flux-model}) into Eq.~(\ref{conduction}) and we solve $\nabla^{2} T$ = 0 by considering Poisson kernel of the disk shape domain. Then, we obtain the temperature distribution,
\begin{equation}
\begin{split}\label{temperature}
T(r, \phi)& =~T_{0} - \frac{\Delta H_{v}Dc_{v}(1-R_{H})}{k_{w}}\biggl[\frac{2E(1-\alpha)^{2}}{\pi} -\alpha \\ 
& + (1-\alpha)\left(\tilde{r}\right)\mbox{cos}\phi + \sum^{\infty}_{n = 1}\left(\tilde{r}\right)^{2n}a_{n}\mbox{cos}(2n\phi) \biggl],
\end{split}
\end{equation}
where $E$ is the elliptic integral of the 2nd kind, $T_{0}$ the temperature at the pinning point ($\phi$ = $\pi$), and $\tilde{r}$ = $r/R$. Fig.~\ref{analytic_results}(a) shows the temperature distribution. The coefficient $a_n$ = $\frac{(1-\alpha)^{2n}}{2^{2n-1}}$ $\binom{1/2}{n}$$F_{1}$[-$\frac{1}{2}$+$n$; $\frac{1}{2}$+$n$; $1+2n$; $(1-\alpha)^2$] from the temperature analytical solution. Here, $\binom{1/2}{n}$ = $\frac{1/2!}{n!(n-1/2)!}$ where $n$ = 0, 1, 2, ..., $\infty$. $F_{1}$ is the Gauss's hypergeometric function.
\begin{figure}
\end{figure}

\subsection{Hydrodynamic model in the polar coordinate}
In this problem, for the analytical model, we consider only the moment that the contact line has the relative receding motion and we assume that it is quasi-steady. Furthermore, a two-dimensional polar coordinate system is considered as shown in Fig.~\ref{schematic}(d). The reason is that the most dominant effect is from the contact line motion. 
In this problem, due to $Re$ $\ll$ 1, we assume that the flow is governed by the Stokes equations in the polar coordinates. To obtain the streamfunction for this problem, we solve the biharmonic equation,
\begin{equation}
\nabla^4\Psi(r, \phi) = 0,
\label{stokes_eq}
\end{equation}
where the general Fourier series solution to biharmonic equation, reads~\cite{michell1899direct}
\begin{equation}
\begin{split}\label{streamfunction}
\Psi(r, \phi)& = A{_{01}}r^2 + A{_{02}}r^2 \mbox{ln}r+A{_{03}} \mbox{ln}r+A{_{04}}\phi\\
& + (A{_{10}}r+A{_{11}}r^{3}+A{_{12}}r\mbox{ln}r + A{_{13}} r\phi + A{_{14}}/r)\mbox{cos}\phi\\
& + (B_{10}r + B_{11}r^{3} +B_{12}r\mbox{ln}r + B_{13}r\theta +B_{14}/r)\mbox{sin}\phi\\
& + \sum^{\infty}_{n = 2}\left( A_{n1}r^{n+2} + A_{n2}r^{n} + A_{n3}/r^{n-2} + A_{n4}/r^{n} \right)\mbox{cos}(n\phi)\\
& + \sum^{\infty}_{n = 2}\left( B_{n1}r^{n+2} + B_{n2}r^{n} + B_{n3}/r^{n-2} + B_{n4}/r^{n} \right)\mbox{sin}(n\phi).
\end{split}
\end{equation}
To obtain the solution, we use a typical three boundary conditions, i.e. (1) the periodicity of the streamfunction and velocity field, (2) no divergence of streamfunction and velocity at the center ($r = 0$) from the physical observation, and (3) the kinematic boundary condition along the contact line $(U_{r}(R, \phi)$ = 0). Additionally, the temperature distribution from energy has to be applied to the hydrodynamic solution. Thus, using the non-uniform evaporation flux approximation, we postulate that the circulating flow is induced by thermal Marangoni stresses along the contact line and therefore the final condition is $|$$\sigma_{r\phi}(R,\phi)$$|$ = $\sigma_{Ma}$. Then, the streamfunction reads
\begin{equation}
\begin{split}
\Psi(r,\phi) & = U_{Ma}R \left[ (1-\alpha)\biggl[ \left(\frac{r}{R}\right)^{3} - \left(\frac{r}{R}\right) \right]\mbox{sin}\phi \\
&+ \sum^{\infty}_{n = 1} a_{n}\left[ \left(\frac{r}{R}\right)^{2n+2} -\left(\frac{r}{R}\right)^{2n} \right]\mbox{sin}(2n\phi) \biggl].
\end{split}
\label{streamfunction_solution}
\end{equation}
where $U_{Ma}$ is the velocity driven by thermal Marangoni effects, which is described as $(|\beta| \Delta H_{v}Dc_{v}[1-R_{H}])/(4\mu k_{w})$. The coefficient $a_{n}$ is $\frac{(1-\alpha)^{2n}}{2^{2n-1}}$$\binom{1/2}{n}$ $F_{1}$ $[ -\frac{1}{2} + n; \frac{1}{2} + n;$ $1 + 2n; (1-\alpha)^{2}]$. The vapor diffusion coefficient $D$ is 2 $\times$ 10$^{-5}$ cm/s, the surface tension-temperature coefficient of water $\beta$ = -1.514 $\times$ 10$^{-4}$ N$\cdot$K$^{-1}$m$^{-1}$, the dynamic viscosity of water $\mu$ = 10$^{-3}$ Pa$\cdot$s, the thermal conductivity of water $k_{w}$ = 0.6 W$\cdot$K$^{-1}$m$^{-1}$, the specific latent heat of evaporation $\Delta$$H_{v}$ = 2.2 $\times$ 10$^{6}$ J$\cdot$kg$^{-1}$, and the saturated concentration of water in the air $c_{v}$ = 6.8 $\times$ 10$^{-3}$ kg$\cdot$m$^{-3}$.

From Eq.~\ref{streamfunction_solution}, we define the velocity components ($U_{r}$, $U_{\phi}$) in terms of streamfunction $\Psi$($r$, $\phi$) as
\begin{equation}\label{velocity}
U_{r}(r, \phi) = \frac{1}{r}\frac{\partial \Psi(r, \phi)}{\partial \phi},~~ U_{\phi}(r, \phi) = - \frac{\partial \Psi(r, \phi)}{\partial r}.
\end{equation}

For the boundary conditions at the liquid-gas interface ($r$ = $R$), we have a zero normal velocity on the liquid interface and there are thermal-Marangoni stresses along the interface, i.e.
\begin{equation}\label{boundary}
\begin{split}
& U_{r}(R, \phi) = 0, \\
& \sigma_{r\phi}(R, \phi) = \mu\left[ -r\frac{\partial}{\partial r}\left(\frac{1}{r}\frac{\partial \Psi}{\partial r}\right) + \frac{1}{r^2}\frac{\partial^2 \Psi}{\partial \phi^2}\right] = \sigma_{Ma}\textbf{t},
\end{split}
\end{equation}
where $\textbf{t}$ is a tangential unit vector. The thermal-Marangoni stress along the liquid-gas interface is $\sigma_{Ma}$ = $\frac{\beta}{R}\frac{\partial T}{\partial \phi}\Bigl\lvert_{r = R}$ where $\beta$ is the surface tension-temperature coefficient of water. Here, the Marangoni number ($Ma$ = -($\beta\Delta$T$t_{f}$)/($\mu$ $R$))~\citep{hu2006marangoni} is approximated as 30 where the temperature difference $(\Delta T = T(R,\pi)-T(R,0))$ is about 0.3 mK (see Fig.~\ref{analytic_results}(a)) and $t_{f}$ is the total drying time, i.e. 1000 s. This temperature difference $\Delta T$ is obtained from the analytical result that is based on the relative evaporation flux along the contact line where $\alpha$ is $r(t)/r(t - \Delta t)$ where $t$ = 186 s and $\Delta t$ = 1 s that is from the experimental observation. Furthermore, we used the boundary condition that is the periodicity of the streamfunction and velocity field, i.e. $\Psi$($r$, $\phi$) = $\Psi$($r$, $\phi+2\pi$). As we mentioned before, we assumed that no divergence of the streamlines and the velocity field at the center of the droplet, which are confirmed from the measurement results.

\begin{figure}
  \centering
      \includegraphics[trim=0.04cm 0.01cm 0cm 0.02cm, clip=true, width=1.0\textwidth]{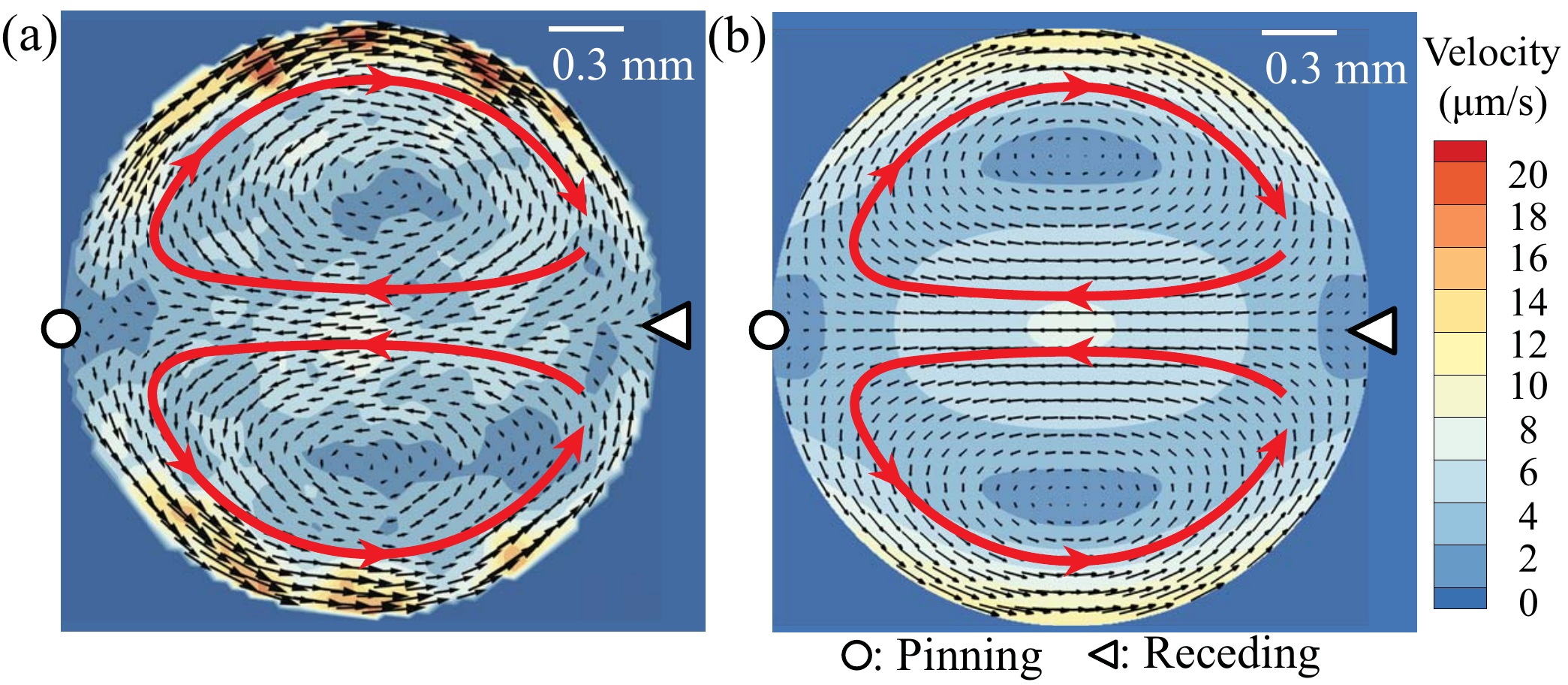}
  \caption{Comparison of the experiment and asymptotic model result; the velocity vector field (black arrows); the velocity magnitude map (color contours); and the circulation direction (red arrows). (a) Tomographic PIV results at $t$ = 186 s close to the substrate (at $z$ = 50 $\upmu$m). (b) Asymptotic solution for the flow field at $t$ = 186 s and $\Delta t$ = 1 s where $\alpha$ = $r$($t$)$/$$r$($t - \Delta t$). $r(t)$ = $\pi^{-1/2}$$\sqrt{(-6.45 \times 10^{-9})t + 7.15 \times 10^{-6}}$ is obtained from experimental results (see Fig.~S3 in the SI). $\circ$ and $\triangleleft$ are the relative pinning and receding site, respectively.}
  \label{comparison}
\end{figure}
Based on the boundary conditions, the analytical solution can be obtained as $\Psi(r,\phi) = U_{Ma}R$ $[ (1-\alpha)[ \tilde{r}^{3} -\tilde{r}]\mbox{sin}\phi$ $+\sum^{\infty}_{n = 1} a_{n}[\tilde{r}^{2n+2} -\tilde{r}^{2n}]\mbox{sin}(2n\phi)]$ (see the streamfunctions in Fig.~\ref{analytic_results}(b)). From Eq.~(\ref{velocity}), the radial and azimuthal velocity fields are obtained as:
\begin{equation}
\begin{split}\label{velocity_R}
U_{r}(r,\phi) = &~U_{Ma} \bigg[ (1-\alpha)\left[ \tilde{r}^{2} - 1 \right]\mbox{cos}\phi \\
&+ \sum^{\infty}_{n = 1} 2n \cdot a_{n}\left[ \tilde{r}^{2n+1} -\tilde{r}^{2n-1} \right]\mbox{cos}(2n\phi) \bigg],
\end{split}
\end{equation}
\begin{equation}
\begin{split}\label{velocity_phi}
U_{\phi}(r,\phi) = &~U_{Ma}\bigg[ (1-\alpha)\left[ \left(1 - 3\tilde{r}\right)^{2}\right]\mbox{sin}\phi \\
&+ \sum^{\infty}_{n = 1} 2\cdot a_{n}\left[ (n+1)\tilde{r}^{2n+1} -n\left(\tilde{r}\right)^{2n-1} \right]\mbox{sin}(2n\phi) \bigg],
\end{split}
\end{equation}
where the velocity driven by thermal Marangoni effects, $U_{Ma}$ = $(|\beta|\Delta H_{v}Dc_{v}[1-RH])/(4\mu k_{w})$. With respect to non-uniform temperature distribution (i.e. Eq.~(\ref{temperature})), the analytical solution of a velocity field is presented in Fig.~\ref{comparison}(a). The magnitude of a maximum velocity is about 12 $\upmu$m/s, which is slightly smaller than the experimental results (see Fig~\ref{comparison}(b)). In this study, for the sake of simplicity, we assumed the contact angle of the droplet is 90$^{\circ}$ although the actual contact angle is about 50$^{\circ}$. Due to this discrepancy, the model underestimates the internal flow speed because the evaporation flux near the contact line depends on the contact angle~\citep{deegan1997capillary, hu2002evaporation}. We should note that the model has a good agreement with the experimental result as shown in Fig.~\ref{comparison}, albeit a small difference between the experimental result and theory. 

\section{Conclusion}
In this paper, we have investigated an evaporating droplet where the contact line receding pattern is non-uniform.
We observe a pair of azimuthal vortices inside an evaporating droplet, which is obtained by tomographic PIV (see Fig.~\ref{3Dflow}).
We find that the internal flow field is related to the contact line receding mode (see Fig.~\ref{flow_evolution}).
The irregularly receding contact line induces some local effect. 
In this study, we model that the non-uniform receding motion could be associated with the non-uniform evaporating flux, which eventually changes the local temperature in the droplet. 
To understand this flow field, we consider a non-uniform evaporation rate along the azimuthal direction.
We should admit that the current analytical model is not perfect, but it predicts well the experimental result, as shown in Fig.~\ref{comparison}.
We believe that the current research is the first study to measure the three-dimensional flow field when the contact line has a stick-and-slip mode during evaporation. For the possible future work, we could suggest that the internal flow field can be measured while the temperature field along the contact line is accurately controlled and the investigation of the various dried patterns by increasing particles concentration while the contact line randomly recedes.

\section{Acknowledgments}
The research leading to these results has received funding from the European Community's Seventh Framework Programme (FP7/2007-2013) under grant agreement No. 215723. We thank S. Armini and T. Delande from IMEC for the preparation of the SAMs samples. We acknowledge useful conversation with B. Wieneke, S. Tokgoez, and G. Elsinga, the support of ASML Netherlands B.V. in funding the PIV set-up and of B. Ramin and M. Riepen from ASML. Furthermore, we thank to O. Shardt, I. Griffiths, S. Wilson, A. Darhuber, and H. Masoud for valuable discussions. This work was partially supported by Basic Science Research Program through the National Research Foundation of Korea funded by the Ministry of Science (NRF-2018R1C1B6004190 and NRF-2019M1A7A1A02089979).

\bibliographystyle{unsrt}
\bibliography{Vortexflow}

\begin{thebibliography}{10}

\bibitem{deegan1997capillary}
Robert~D Deegan, Olgica Bakajin, Todd~F Dupont, Greb Huber, Sidney~R Nagel, and
  Thomas~A Witten.
\newblock Capillary flow as the cause of ring stains from dried liquid drops.
\newblock {\em \textit{Nature}}, 389(6653):827--829, 1997.

\bibitem{deegan2000contact}
Robert~D Deegan, Olgica Bakajin, Todd~F Dupont, Greg Huber, Sidney~R Nagel, and
  Thomas~A Witten.
\newblock Contact line deposits in an evaporating drop.
\newblock {\em \textit{Phys. Rev. E}}, 62(1):756, 2000.

\bibitem{kim2016controlled}
Hyoungsoo Kim, Fran{\c{c}}ois Boulogne, Eujin Um, Ian Jacobi, Ernie Button, and
  Howard~A Stone.
\newblock Controlled uniform coating from the interplay of marangoni flows and
  surface-adsorbed macromolecules.
\newblock {\em \textit{Phys. Rev. Lett.}}, 116(12):124501, 2016.

\bibitem{rothschild2004liquid}
M~Rothschild, TM~Bloomstein, RR~Kunz, V~Liberman, M~Switkes, ST~Palmacci, JHC
  Sedlacek, D~Hardy, and A~Grenville.
\newblock Liquid immersion lithography: Why, how, and when?
\newblock {\em \textit{J. Vac. Sci. Technol., B}}, 22(6):2877--2881, 2004.

\bibitem{calvert2001inkjet}
Paul Calvert.
\newblock Inkjet printing for materials and devices.
\newblock {\em \textit{Chem. Mater.}}, 13(10):3299--3305, 2001.

\bibitem{kuang2014controllable}
Minxuan Kuang, Libin Wang, and Yanlin Song.
\newblock Controllable printing droplets for high-resolution patterns.
\newblock {\em \textit{Adv. Mater.}}, 26(40):6950--6958, 2014.

\bibitem{deegan2000pattern}
Robert~D Deegan.
\newblock Pattern formation in drying drops.
\newblock {\em \textit{Phys. Rev. E}}, 61(1):475, 2000.

\bibitem{park2006control}
Jungho Park and Jooho Moon.
\newblock Control of colloidal particle deposit patterns within picoliter
  droplets ejected by ink-jet printing.
\newblock {\em \textit{Langmuir}}, 22(8):3506--3513, 2006.

\bibitem{lee2018uniform}
Seung~Yeol Lee, Hyoungsoo Kim, Shin-Hyun Kim, and Howard~A Stone.
\newblock Uniform coating of self-assembled noniridescent colloidal
  nanostructures using the marangoni effect and polymers.
\newblock {\em \textit{Phys. Rev. Appl.}}, 10(5):054003, 2018.

\bibitem{lenshof2009acoustic}
Andreas Lenshof, Asilah Ahmad-Tajudin, Kerstin J$\ddot{a}$r{\aa}s, Ann-Margret
  Sw$\ddot{a}$rd-Nilsson, Lena {\AA}berg, Gy$\ddot{o}$rgy Marko-Varga, Johan
  Malm, Hans Lilja, and Thomas Laurell.
\newblock Acoustic whole blood plasmapheresis chip for prostate specific
  antigen microarray diagnostics.
\newblock {\em \textit{Anal. Chem.}}, 81(15):6030--6037, 2009.

\bibitem{dugas2005droplet}
Vincent Dugas, J{\'e}r{\^o}me Broutin, and Eliane Souteyrand.
\newblock Droplet evaporation study applied to dna chip manufacturing.
\newblock {\em \textit{Langmuir}}, 21(20):9130--9136, 2005.

\bibitem{martusevich2007morphology}
Andrew~Kimovich Martusevich, Yury Zimin, and Anna Bochkareva.
\newblock Morphology of dried blood serum specimens of viral hepatitis.
\newblock {\em \textit{Hepat. Mon.}}, 7:207--210, 2007.

\bibitem{brutin2011pattern}
David Brutin, Benjamin Sobac, Boris Loquet, and Jos{\'e} Sampol.
\newblock Pattern formation in drying drops of blood.
\newblock {\em \textit{J. Fluid Mech.}}, 667:85--95, 2011.

\bibitem{hu2002evaporation}
Hua Hu and Ronald~G Larson.
\newblock Evaporation of a sessile droplet on a substrate.
\newblock {\em \textit{J. Phys. Chem. B}}, 106(6):1334--1344, 2002.

\bibitem{hu2005analysis}
Hua Hu and Ronald~G Larson.
\newblock Analysis of the microfluid flow in an evaporating sessile droplet.
\newblock {\em \textit{Langmuir}}, 21(9):3963--3971, 2005.

\bibitem{masoud2009analytical}
Hassan Masoud and James~D Felske.
\newblock Analytical solution for stokes flow inside an evaporating sessile
  drop: Spherical and cylindrical cap shapes.
\newblock {\em \textit{Phys. Fluids}}, 21:042102, 2009.

\bibitem{petsi2008stokes}
AJ~Petsi and VN~Burganos.
\newblock Stokes flow inside an evaporating liquid line for any contact angle.
\newblock {\em \textit{Phys. Rev. E}}, 78(3):036324, 2008.

\bibitem{marin2011order}
{\'A}lvaro~G Mar{\'\i}n, Hanneke Gelderblom, Detlef Lohse, and Jacco~H
  Snoeijer.
\newblock Order-to-disorder transition in ring-shaped colloidal stains.
\newblock {\em \textit{Phys. Rev. Lett.}}, 107(8):085502, 2011.

\bibitem{popov2005evaporative}
Yuri~O Popov.
\newblock Evaporative deposition patterns: Spatial dimensions of the deposit.
\newblock {\em \textit{Phys. Rev. E}}, 71(3):036313, 2005.

\bibitem{park2019control}
Jonghyeok Park, Junil Ryu, Hyung~Jin Sung, and Hyoungsoo Kim.
\newblock Control of solutal marangoni-driven vortical flows and enhancement of
  mixing efficiency.
\newblock {\em \textit{J. Colloid Interface Sci.}}, 561(1):408--415, 2020.

\bibitem{sefiane2013thermal}
Khellil Sefiane, Yuki Fukatani, Yasuyuki Takata, and Jungho Kim.
\newblock Thermal patterns and hydrothermal waves (htws) in volatile drops.
\newblock {\em \textit{Langmuir}}, 29(31):9750--9760, 2013.

\bibitem{de2014thermocapillary}
Raf De~Dier, Wouter Sempels, Johan Hofkens, and Jan Vermant.
\newblock Thermocapillary fingering in surfactant-laden water droplets.
\newblock {\em \textit{Langmuir}}, 30(44):13338--13344, 2014.

\bibitem{karapetsas2012convective}
George Karapetsas, Omar~K Matar, Prashant Valluri, and Khellil Sefiane.
\newblock Convective rolls and hydrothermal waves in evaporating sessile drops.
\newblock {\em \textit{Langmuir}}, 28(31):11433--11439, 2012.

\bibitem{thiele2006depinning}
Uwe Thiele and Edgar Knobloch.
\newblock On the depinning of a driven drop on a heterogeneous substrate.
\newblock {\em \textit{New J. Phys.}}, 8(12):313, 2006.

\bibitem{park2012change}
Jun~Kwon Park, Jeongeun Ryu, Bonchull~C Koo, Sanghyun Lee, and Kwan~Hyoung
  Kang.
\newblock How the change of contact angle occurs for an evaporating droplet:
  Effect of impurity and attached water films.
\newblock {\em \textit{Soft Matter}}, 8(47):11889--11896, 2012.

\bibitem{bhardwaj2009pattern}
Rajneesh Bhardwaj, Xiaohua Fang, and Daniel Attinger.
\newblock Pattern formation during the evaporation of a colloidal nanoliter
  drop: {A} numerical and experimental study.
\newblock {\em \textit{New J. Phys.}}, 11(7):075020, 2009.

\bibitem{orejon2011stick}
Daniel Orejon, Khellil Sefiane, and Martin~ER Shanahan.
\newblock {S}tick--{S}lip of evaporating droplets: Substrate hydrophobicity and
  nanoparticle concentration.
\newblock {\em \textit{Langmuir}}, 27(21):12834--12843, 2011.

\bibitem{shanahan1995simple}
Martin~ER Shanahan.
\newblock Simple theory of ``stick-slip" wetting hysteresis.
\newblock {\em \textit{Langmuir}}, 11(3):1041--1043, 1995.

\bibitem{saenz2015evaporation}
P.~J. S{\'a}enz, K.~Sefiane, J.~Kim, O.~K. Matar, and P.~Valluri.
\newblock Evaporation of sessile drops: a three-dimensional approach.
\newblock {\em \textit{J. Fluid Mech.}}, 772:705--739, 2015.

\bibitem{kim2016tomographic}
Hyoungsoo Kim, Naser Belmiloud, and Paul~W Mertens.
\newblock Tomographic piv measurement of internal complex flow of an
  evaporating droplet with non-uniformly receding contact lines.
\newblock {\em \textit{J. Kor. Soc. Vis.}}, 14(2):31--39, 2016.

\bibitem{howarter2006optimization}
John~A Howarter and Jeffrey~P Youngblood.
\newblock Optimization of silica silanization by 3-aminopropyltriethoxysilane.
\newblock {\em \textit{Langmuir}}, 22(26):11142--11147, 2006.

\bibitem{howarter2007surface}
John~A Howarter and Jeffrey~P Youngblood.
\newblock Surface modification of polymers with 3-aminopropyltriethoxysilane as
  a general pretreatment for controlled wettability.
\newblock {\em \textit{Macromolecules}}, 40(4):1128--1132, 2007.

\bibitem{schwartz2001mechanisms}
Daniel~K Schwartz.
\newblock Mechanisms and kinetics of self-assembled monolayer formation.
\newblock {\em \textit{Annu. Rev. Phys. Chem.}}, 52(1):107--137, 2001.

\bibitem{zhao1996mechanism}
Xiaolin Zhao and Raoul Kopelman.
\newblock Mechanism of organosilane self-assembled monolayer formation on
  silica studied by second-harmonic generation.
\newblock {\em \textit{J. Phys. Chem. A}}, 100(26):11014--11018, 1996.

\bibitem{iimura2000study}
Ken-Ichi Iimura, Yayoi Nakajima, and Teiji Kato.
\newblock A study on structures and formation mechanisms of self-assembled
  monolayers of $n$-alkyltrichlorosilanes using infrared spectroscopy and
  atomic force microscopy.
\newblock {\em \textit{Thin Solid Films}}, 379(1):230--239, 2000.

\bibitem{fellah2012facilitating}
Abdenor Fellah, Naser Belmiloud, Richard~G Haverkamp, Yacine Hemar, Don Otter,
  and Martin~AK Williams.
\newblock Facilitating high-force single-polysaccharide stretching using
  covalent attachment of one end of the chain.
\newblock {\em \textit{Carbohydr. Polym.}}, 87(1):806--815, 2012.

\bibitem{adrian2010particle}
Ronald~J Adrian and Jerry Westerweel.
\newblock {\em \textit{Particle Image Velocimetry}}.
\newblock Cambridge University Press, 2010.

\bibitem{wieneke2008volume}
B~Wieneke.
\newblock Volume self-calibration for {3D} particle image velocimetry.
\newblock {\em \textit{Exp. Fluids}}, 45(4):549--556, 2008.

\bibitem{kim2013comparison}
Hyoungsoo Kim, Jerry Westerweel, and Gerrit~E Elsinga.
\newblock Comparison of {Tomo}-{PIV} and 3{D}-{PTV} for microfluidic flows.
\newblock {\em \textit{Meas. Sci. Technol.}}, 24(2):024007, 2013.

\bibitem{elsinga2008tomographic}
Gerrit~Einte Elsinga.
\newblock \textit{Tomographic particle image velocimetry and its application to
  turbulent boundary layers}.
\newblock {\em PhD dissertation, Delft University of Technology, Delft
  (http://repository.tudelft.nl/file/1003861/379883)}, 2008.

\bibitem{kim2011full}
Hyoungsoo Kim, Sebastian Gro{\ss}e, Gerrit~E Elsinga, and Jerry Westerweel.
\newblock Full 3{D}-3{C} velocity measurement inside a liquid immersion
  droplet.
\newblock {\em \textit{Exp. Fluids}}, 51(2):395--405, 2011.

\bibitem{lorenzini2013thermal}
Giulio Lorenzini and Onorio Saro.
\newblock Thermal fluid dynamic modelling of a water droplet evaporating in
  air.
\newblock {\em \textit{Int. J. Heat Mass Transfer}}, 62:323--335, 2013.

\bibitem{michell1899direct}
J~H Michell.
\newblock On the direct determination of stress in an elastic solid, with
  application to the theory of plates.
\newblock {\em \textit{Proc. London Math. Soc.}}, 1(1):100--124, 1899.

\bibitem{hu2006marangoni}
Hua Hu and Ronald~G Larson.
\newblock Marangoni effect reverses coffee-ring depositions.
\newblock {\em \textit{J. Phys. Chem. B}}, 110(14):7090--7094, 2006.

\end{thebibliography}

\end{document}